# Universal spectrum for DNA base C+G concentration variability in Human chromosome Y

## A. M. Selvam[1]

Deputy Director (Retired)

Indian Institute of Tropical Meteorology, Pune 411 008, India

Email: amselvam@eth.net

Web sites: http://www.geocities.com/amselvam

http://amselvam.tripod.com/index.html

## Abstract

The spatial distribution of DNA base sequence A, C, G and T exhibit selfsimilar fractal fluctuations and the corresponding power spectra follow inverse power law of the form $1/f^{\alpha}$ where $f$ is the frequency and $\alpha$ the exponent. Inverse power law form for power spectra implies the following: (1) A scale invariant eddy continuum, namely, the amplitudes of component eddies are related to each other by a scale factor alone. In general, the scale factor is different for different scale ranges and indicates a multifractal structure for the spatial distribution of DNA base sequence. (2) Scale invariance of eddies also implies long-range spatial correlations of the eddy fluctuations. Multifractal structure to space-time fluctuations and the associated inverse power law form for power spectra is generic to spatially extended dynamical systems in nature and is a signature of self-organized criticality. Mathematical models for the simulation and prediction of fractal fluctuations of dynamical systems are nonlinear and do not have analytical solutions. Finite precision computer realizations of non-linear mathematical models of dynamical systems also exhibit self-organized criticality manifested as sensitive dependence on initial conditions and give chaotic solutions resulting in 'deterministic chaos'. The exact physical mechanism for the observed self-organized criticality is not yet identified. The author has developed a general systems theory where quantum mechanical laws emerge as self-consistent explanations for the observed long-range space-time correlations in macro-scale dynamical systems, i.e., the apparently chaotic fractal fluctuations are signatures of quantum-like chaos in dynamical systems. The model also provides unique quantification for the observed inverse power law form for power spectra in terms of the statistical normal distribution. In this paper it is shown that the frequency distribution of the bases C+G in all available contiguous sequences for Human chromosome Y DNA exhibit model predicted quantum-like chaos.

## 1. Introduction

Long-range space-time correlations, manifested as the selfsimilar fractal geometry to the spatial pattern, concomitant with inverse power law form for power spectra of space-time fluctuations are

_______________________

Present corresponding address:
B1 Aradhana, 42/2A Shivajinagar,
Pune 411 005, India[1]

generic to spatially extended dynamical systems in nature and are identified as signatures of self-organized criticality. A representative example is the selfsimilar fractal geometry of His-Purkinje system whose electrical impulses govern the interbeat interval of the heart. The spectrum of interbeat intervals exhibits a broadband inverse power law form $f^{\alpha}$ where $f$ is the frequency and $\alpha$ the exponent. Self-organized criticality implies non-local connections in space and time, i.e., long-term memory of short-term spatial fluctuations in the extended dynamical system that acts as a unified whole communicating network.

Finite precision computer realizations of nonlinear mathematical models of real world dynamical systems also exhibit self-organized criticality manifested as sensitive dependence on initial conditions and identified as deterministic chaos. The author has developed a general systems theory that predicts the observed self-organized criticality in model and real world dynamical systems as intrinsic to quantumlike chaos governing the pattern evolution in space and time. A summary of the model is presented with applications to biological systems.

'Nonlinear dynamics and chaos', a multidisciplinary area of intensive research in recent years (since 1980s) has helped identify universal characteristics of spatial patterns (forms) and temporal fluctuations (functions) of disparate dynamical systems in nature. Examples of dynamical systems, i.e., systems which change with time include biological (living) neural networks of the human brain which responds as a unified whole to a multitude of input signals and the non-biological (non-living) atmospheric flow structure which exhibits teleconnections, i.e., long-range space-time correlations. Spatially extended dynamical systems in nature exhibit selfsimilar fractal geometry to the spatial pattern. The sub-units of selfsimilar structures resemble the whole in shape. The name 'fractal' coined by Mandelbrot (1977) indicates non-Euclidean or fractured (broken) Euclidean structures. Traditional Euclidean geometry discusses only three-, two- and one-dimensional objects, representative examples being sphere, rectangle and straight line respectively. Objects in nature have irregular non-Euclidean shapes, now identified as fractals and the fractal dimension $D$ is given as $D=dlogM/dlogR$, where $M$ is the mass contained within a distance $R$ from a point within the extended object. A constant value for $D$ indicates uniform stretching on logarithmic scale. Objects in nature, in general exhibit multifractal structure, i.e., the fractal dimension $D$ varies with length scale $R$. The fractal structure of physiological systems has been identified (Goldberger et al., 1990, 2002; West, 1990, 2004). The global atmospheric cloud cover pattern also exhibits selfsimilar fractal geometry (Lovejoy and Schertzer, 1986). Selfsimilarity implies long-range spatial correlations, i.e., the larger scale is a magnified version of the smaller scale with enhancement of fine structure. Selfsimilar fractal structures in nature support functions, which fluctuate on all scales of time. For example, the neural network of the human brain responds to a multitude of sensory inputs on all scales of time with long-term memory update and retrieval for appropriate global response to local input signals. Fractal architecture to the spatial pattern enables integration of a multitude of signals of all space-time scales so that the dynamical system responds as a unified whole to local stimuli, i.e., short term fluctuations are carried as internal structure to long-term fluctuations. Fractal networks therefore function as dynamic memory storage devices, which integrate short-term fluctuations into long-period fluctuations. The irregular (nonlinear) variations of fluctuations in dynamical systems are therefore broadband because of coexistence of fluctuations of all scales. Power spectral analysis (MacDonald, 1989) is conventionally used to resolve the periodicities (frequencies) and their amplitudes in time series data of fluctuations. The power spectrum is plotted on log-log scale as the intensity represented by variance (amplitude squared) versus the period (frequency) of the component periodicities. Dynamical systems in nature exhibit inverse power law form $1/f^\alpha$ where $f$ is the frequency (1/period) and $\alpha$ the exponent for the power spectra of space-time fluctuations indicating selfsimilar fluctuations on all space-time scales, i.e., long-range space-time correlations. The amplitudes of short-term and long-term are related by a scale factor alone, i.e., the space-time fluctuations exhibit scale invariance or long-range space-time correlations, which are independent of the exact details of dynamical mechanisms underlying the fluctuations at different scales. The universal characteristics of spatially extended dynamical systems, namely, the fractal structure to the space-time fluctuation pattern and inverse power law form for power spectra of space-time fluctuations are identified as signatures of self-organized criticality (Bak et al., 1988). Self-organized criticality implies non-local connections in space and time in real world dynamical systems.

Surprisingly, such long-range space-time correlations had been earlier identified (Gleick, 1987) as sensitive dependence on initial conditions of finite precision computer realizations of nonlinear mathematical models of dynamical systems and named 'deterministic chaos'. Deterministic chaos is therefore a signature of self-organized criticality in computed model solutions.

It has not been possible to identify the exact mechanism underlying the observed universal long-range space-time correlations in natural dynamical systems and in computed solutions of model dynamical systems. The physical mechanisms responsible for self-organized criticality should be independent of the exact details (physical, chemical, physiological, biological, computational system etc.) of the dynamical system so as to be universally applicable to all dynamical systems (real and model).

Atmospheric flows exhibit self-organized criticality as manifested in the fractal geometry to the global cloud cover pattern concomitant with inverse power law form for power spectra of temporal fluctuations documented and discussed in detail by Tessier et al. (1993). Standard models for atmospheric flow dynamics cannot explain the observed self-organized criticality in atmospheric flows satisfactorily. The author has developed a general systems theory for atmospheric flows (Selvam, 1990; Selvam and Fadnavis, 1998 and all references therein) that predicts the observed self-organized criticality as intrinsic to quantumlike chaos governing flow dynamics. The model concepts have also been applied to show that deterministic chaos in computed solutions of model dynamical systems is a direct consequence of roundoff error in finite precision iterative computations (Selvam, 1993) incorporated in long-term numerical integration schemes used for numerical solutions.

In the following Section 2, the model for self-organized criticality in atmospheric flows is first summarized and model concepts are shown to be applicable to all real world and model dynamical systems. The concept of self-organized criticality and quantumlike chaos in biological and physiological systems in particular are discussed.

## 2. General systems theory concepts

In summary (Selvam, 1990; Selvam and Fadnavis, 1998), the model is based on Townsend's concept (Townsend, 1956) that large eddy structures form in turbulent flows as envelopes of enclosed turbulent eddies. Such a simple concept that space-time averaging of small-scale structures gives rise to large-scale space-time fluctuations leads to the following important model predictions.

### 2.1 Quantumlike chaos in turbulent fluid flows

Since the large eddy is but the integrated mean of enclosed turbulent eddies, the eddy energy (kinetic) distribution follows statistical normal distribution according to the Central Limit Theorem (Ruhla, 1992). Such a result, that the additive amplitudes of the eddies, when squared, represent probability distributions is found in the subatomic dynamics of quantum systems such as the electron or photon. Atmospheric flows, or, in general turbulent fluid flows follow quantumlike chaos.

### 2.2 Dynamic memory (information) circulation network

The root mean square (r.m.s.) circulation speeds $W$ and $w_*$ of large and turbulent eddies of respective radii $R$ and $r$ are related as

$$W^2 = \frac{2}{\pi} \frac{r}{R} w_*^2 \qquad (1)$$

Eq.(1) is a statement of the law of conservation of energy for eddy growth in fluid flows and implies a two-way ordered energy flow between the larger and smaller scales. Microscopic scale perturbations are carried permanently as internal circulations of progressively larger eddies. Fluid flows therefore act as dynamic memory circulation networks with intrinsic long-term memory of short-term fluctuations. Such "memory of water" is reported by Davenas et al. (1988).

## 2.3 Quasicrystalline structure

The flow structure consists of an overall logarithmic spiral trajectory with Fibonacci winding number and quasiperiodic Penrose tiling pattern for internal structure (Fig.1). Primary perturbation $OR_O$ (Fig.1) of time period $T$ generates return circulation $OR_1R_O$ which, in turn, generates successively larger circulations $OR_1R_2$, $OR_2R_3$, $OR_3R_4$, $OR_4R_5$, etc., such that the successive radii form the Fibonacci mathematical number series, i.e., $OR_1/OR_O = OR_2/OR_1 = \ldots\ldots = \tau$ where $\tau$ is the golden mean equal to $(1+\sqrt{5})/2 \approx 1.618$. The flow structure therefore consists of a nested continuum of vortices, i.e., vortices within vortices.

Figure 1: The quasiperiodic Penrose tiling pattern which forms the internal structure at large eddy circulations

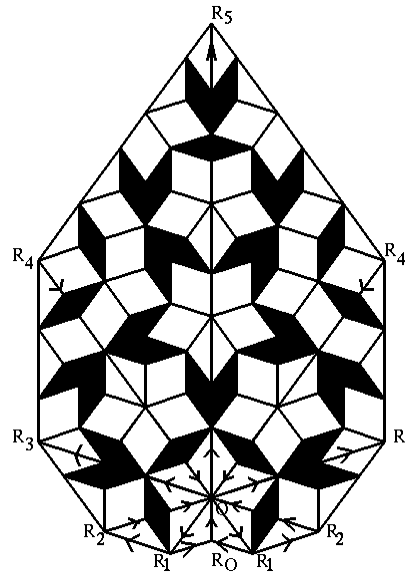

The quasiperiodic Penrose tiling pattern with five-fold symmetry has been identified as quasicrystalline structure in condensed matter physics (Janssen, 1988). The self-organized large eddy growth dynamics, therefore, spontaneously generates an internal structure with the five-fold symmetry of the dodecahedron, which is referred to as the icosahedral symmetry, e.g., the geodesic dome devised by Buckminster Fuller. Incidentally, the pentagonal dodecahedron is, after the helix, nature's second favourite structure (Stevens, 1974). Recently the carbon macromolecule $C_{60}$, formed by condensation from a carbon vapour jet, was found to exhibit the icosahedral symmetry of the closed soccer ball and has been named Buckminsterfullerene or footballene (Curl and Smalley, 1991). Self-organized quasicrystalline pattern formation therefore exists at the molecular level also and may result in condensation of specific biochemical structures in biological media. Logarithmic spiral formation with Fibonacci winding number and five-fold symmetry possess maximum packing efficiency for component parts and are manifested strikingly in Phyllotaxis (Jean, 1992a,b; 1994) and is common to nature (Stevens, 1974; Tarasov, 1986).

Model predicted spiral flow structure is seen vividly in the hurricane cloud cover pattern. Spiral waves are observed in many dynamical systems. Examples include Belousov-Zhabotinksy chemical reaction and also in the electrical activity of heart (Steinbock et al., 1993).

## 2.4 Dominant periodicities

Dominant quasi-periodicities $P_n$ corresponding to the internal circulations (Fig.1) $OR_O R_1$, $OR_1 R_2$, $OR_2 R_3$, ….. are given as

$$P_n = T(2 + \tau)\tau^n \tag{2}$$

The dominant quasi-periodicities are equal to *2.2T*, *3.6T*, *5.8T*, *9.5T*, ……for values of *n = -1, 0, 1, 2,…*, respectively (Eq.2). Space-time integration of turbulent fluctuations results in robust broadband dominant periodicities which are functions of the primary perturbation time period *T* alone and are independent of exact details (chemical, electrical, physical etc.) of turbulent fluctuations. Also, such global scale oscillations in the unified network are not affected appreciably by failure of localized microscale circulation networks.

Wavelengths (or periodicities) close to the model predicted values have been reported in weather and climate variability (Selvam and Fadnavis, 1998), prime number distribution (Selvam, 2001a), Riemann zeta zeros (non-trivial) distribution (Selvam, 2001b), Drosophila DNA base sequence (Selvam, 2002), stock market economics (Selvam, 2003), Human chromosome 1 DNA base sequence (Selvam, 2004).

Similar unified communication networks may be involved in biological and physiological systems such as the brain and heart, which continue to perform overall functions satisfactorily in spite of localized physical damage. Structurally stable network configurations increase insensitivity to parameter changes, noise and minor mutations (Kitano, 2002).

Model predicted dominant quasiperiodicities (years) equal to *2.2*, *3.6*, *5.8*, *9.5*, (Eq.2) generated by the annual cycle (*T*=1 year in Eq.2) of solar heating in atmospheric flows have been identified in global atmospheric weather patterns (Burroughs, 1992) as the quasibiennial oscillation or QBO (*2.2* years), the high frequency (3-4 years) and low frequency (*5.8* years) components of the 3-7 years El Nino-Southern Oscillation (ENSO) cycle and decadic scale (>*9* years) fluctuations. The ENSO cycle in particular is characterized by devastating regional changes in global climate pattern (Philander, 1990) and is now of public concern.

Persistent periodic energy pumping at fixed time intervals (period) $T$ in a fluid medium generates self-sustaining continuum of eddies and results in apparent nonlinear chaotic fluctuations in the fluid medium. Such chaotic optical (laser) emissions are triggered in nonlinear optical medium using a laser energy pump (Harrison and Biswas, 1986). Self-organized broadband structures may therefore be generated in electromagnetic fields also.

Macroscale coherent structures emerge by space-time integration of microscopic domain fluctuations in fluid flows. Such a concept of the autonomous growth of atmospheric eddy continuum with ordered energy flow between the scales is analogous to Prigogine's (Prigogine and Stengers, 1988) concept of the spontaneous emergence of order and organization out of apparent disorder and chaos through a process of self-organization.

### 2.4.1 Emergence of order and coherence in biology

The problem of emergence of macroscopic variables out of microscopic dynamics is of crucial relevance in biology (Vitiello, 1992). In atmospheric flows turbulent fluctuations self-organize to form large eddies which give rise to cloud formations in updraft regions where moisture condenses. Similarly, in biological systems collective microscopic scale behaviour, e.g., self-organization of local information flow in neural networks may initiate global response in the human brain. Biological systems rely on a combination of network and the specific elements involved (Kitano, 2002). The notion that membership in a network could confer stability emerged from Ludwig von Bertalanffy's description of general systems theory in the 1930s and Norbert Wieners description of cybernetics in the 1940s. General systems theory focused in part on the notion of flow, postulating the existence and significance of flow equilibria. In contrast to Cannon's concept that mechanisms should yield homeostasis, general systems theory invited biologists to consider an alternative model of homeodynamics in which nonlinear, non-equilibrium processes could provide stability, if not constancy (Buchman, 2002).

The cell dynamical system model for coherent pattern formation in turbulent flows summarized earlier (Section 2) may provide a general systems theory for biological complexity. General systems theory is a logical-mathematical field, the subject matter of which is the formulation and deduction of those principles which are valid for 'systems' in general, whatever the nature of their component elements or the relations or 'forces' between them (Bertalanffy, 1968; Peacocke, 1989; Klir, 1993).

More than 25 years ago Frohlich (1968, 1970, 1975, 1980) introduced the concept of cooperative vibrational modes between proteins in biological cells. Coherent oscillations in the range of $10^{10}$-$10^{12}$ Hz involving cell membranes, DNA and cellular proteins could be generated by interaction between vibrating electric dipoles contained in the proteins as a result of nonlinear properties of the system. Through long-range effects proper to Frohlich nonlinear electrodynamics a temporospatial link, is in fact, established between all molecules constituting the system. Single molecules may thus act in a synchronized fashion and can no longer be considered as individual. New unexpected features arise from such a dynamic system, reacting as a unified whole entity (Insinnia, 1992). Coherent Frohlich oscillations may be associated with the dynamical pattern formation of intracellular cytoskeletal architecture consisting of networks of filamentous protein polymers, which coordinate and integrate information flow in the biological cell (Dayhoff et al., 1994; Hameroff et al., 1984, 1986,1989; Hotani et al., 1992). Grundler and Kaiser (1992), Kaiser (1992), Tabony and Job (1992) have also discussed biological autoorganization and pattern formation in the context of such coherent oscillations.

## 2.5 Long-range spatiotemporal correlations (coherence)

The logarithmic spiral flow pattern enclosing the vortices $OR_OR_1$, $OR_1R_2$, … may be visualized as a continuous smooth rotation of the phase angle $\theta$ ($R_OOR_1$, $R_OOR_2$, … etc.) with increase in period. The phase angle $\theta$ for each stage of growth is equal to $r/R$ and is proportional to the variance $W^2$ (Eq.1), the variance representing the intensity of fluctuations.

The phase angle gives a measure of coherence or correlation in space-time fluctuations. The model predicted continuous smooth rotation of phase angle with increase in period length associated with logarithmic spiral flow structure is analogous to Berry's phase (Berry, 1988; Kepler et al., 1991) in quantum systems.

## 2.6 Universal spectrum of fluctuations

Conventional power spectral analysis will resolve such a logarithmic spiral flow trajectory as a continuum of eddies (broadband spectrum) with a progressive increase in phase angle.

The power spectrum, plotted on log-log scale as variance versus frequency (period) will represent the probability density corresponding to normalized standard deviation $t$ equal to ($\log L/\log T_{50}$) $-1$ where $L$ is the period in years and $T_{50}$ is the period up to which the cumulative percentage contribution to total variance is equal to $50$. The above expression for normalized standard deviation $t$ follows from model prediction of logarithmic spiral flow structure and model concept of successive growth structures by space-time averaging. Fluctuations of all scales therefore self-organize to form the universal inverse power law form of the statistical normal distribution. Since the phase angle $\theta$ equal to $r/R$ represents the variance $W^2$ (Section 2.5, Eq.1), the phase spectrum plotted similar to variance spectrum will also follow the statistical normal distribution.

Signatures of quantumlike chaos, namely universal inverse power law form for atmospheric eddy energy spectrum and also model predicted quasiperiodicities associated with quasicrystalline Penrose tiling pattern for internal flow structure (Fig.1) have been identified in meteorological parameters (Selvam and Fadnavis, 1998).

## 2.7 Quantum mechanics for subatomic dynamics: apparent paradoxes

The following apparent paradoxes found in the subatomic dynamics of quantum systems (Maddox, 1988) are consistent in the context of atmospheric flows as explained in the following.

### 2.7.1 Wave-particle duality

A quantum system behaves as a wave on some occasions and as a particle at other times. Wave-particle duality is consistent in the context of atmospheric waves, which generate particle-like clouds in a row because of formation of clouds in updrafts and dissipation of clouds in adjacent downdrafts characterizing wave motion (Fig.2).

Figure 2: Illustration of wave-particle duality as physically consistent for quantumlike mechanics in atmospheric flows. Particlelike clouds form in a row because of condensation of water vapour in updrafts and evaporation of condensed water in adjacent downdrafts associated with eddy circulations in atmospheric flows. Wave-particle duality in macroscale real world dynamical systems may be associated with bimodal (formation and dissipation) phenomenological form for manifestation of energy associated with bidirectional energy flow intrinsic to eddy (wave) circulations in the medium of propagation.

## WAVE - PARTICLE DUALITY

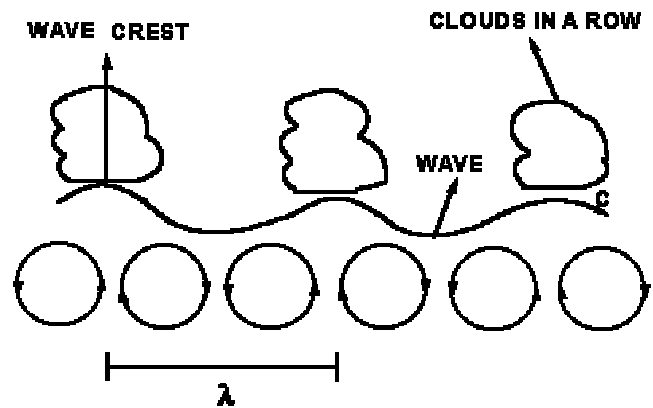

λ = WAVE LENGTH    C = VELOCITY OF PROPAGATION

### 2.7.2 Non-local connection

The separated parts of a quantum system respond as a unified whole to local perturbations. Non-local connection is implicit to atmospheric flow structure quantified in Eq.(1) as ordered two-way energy flow between larger and smaller scales and seen as long-range space-time correlations, namely self-organized criticality. Atmospheric flows self-organize to form a unified network with the quasiperiodic Penrose tiling pattern for internal structure (Fig.1), which provide long-range (non-local) space-time connections.

## 3. Self-organized criticality and quantum-like chaos in computed model dynamical systems

### 3.1 Deterministic chaos

Traditional deterministic mathematical models of dynamical systems based on Newtonian continuum dynamics are nonlinear and do not have analytical solutions. Finite precision computer realizations of nonlinear model dynamical systems are sensitively dependent on initial conditions and give chaotic solutions. Computed solutions therefore exhibit 'deterministic chaos' since deterministic equations give chaotic solutions. Such deterministic chaos was identified nearly a century ago by Poincare (1892) in his study of the three-body problem. Availability of computers with graphical display facilities in late 1950s facilitated numerical solutions and in 1963 Lorenz (1963) identified deterministic chaos in a simple model of atmospheric flows. Ruelle and Takens (1971) were the first to identify deterministic chaos as similar to turbulence in fluid flows. The computed trajectory traces the selfsimilar fractal pattern of the 'strange attractor' so named because of its strange convoluted shape being the final destination of all possible trajectories. 'Chaos Science' is now (since 1980s) an area of intensive research in all branches of science and other areas of human interest (Gleick, 1987). The physics of deterministic chaos is not yet identified. Deterministic chaos is a direct consequence of numerical solutions of error sensitive dynamical systems such as $X_{n+1}=F(X_n)$ where $X_{n+1}$, the value of the variable $X$ at the $(n+1)^{th}$ instant is a function $F$ of $X_n$. Error-feedback loop inherent to such iterative computations magnify exponentially with time the following errors inherent to numerical computations: (1) The continuous dynamical system is computed as a discrete dynamical system because of discretization of space and time in numerical computations with implicit assumption of sub-grid scale homogeneity. (2) Binary computer arithmetic precludes exact number representation at the data input stage itself. (3) Model approximations and assumptions. (4) Roundoff error of finite precision computer arithmetic magnifies exponentially with time the above errors and gives chaotic solutions in iterative computations such as that used in long-term numerical integration schemes in numerical solutions.

Sensitive dependence on initial conditions of computed solutions indicates long-range space-time correlations and is a signature of self-organized criticality as explained earlier.

## 3.2 Universal quantification for deterministic chaos in dynamical systems

Selvam (1993) has shown that round-off error in finite precision computations is analogous to yardstick-length in length measurements. The computed domain at any stage of computation may be resolved as the product $WR$ of the number of units of computation $W$ of yardstick-length $R$ and $wr$ represents the initial uncertainty domain where $w$ is the number of units of computation of precision $r$ to begin with. Iterative computations may be visualized as spatial integration of enclosed higher precision uncertainty domain $wr$ resulting in the larger uncertainty domain $WR$. The above concept is similar to the growth of large eddy structures from turbulent fluctuations. The concepts of cell dynamical system model for growth of large eddy structures in turbulent flows may therefore be applied for the growth of selfsimilar structures in the computed domain. The computed domain when resolved as a function of computational precision is shown (Selvam, 1993) to have an overall logarithmic spiral envelope with the quasiperiodic Penrose tiling pattern for the internal structure.

The computed dynamical system follows quantumlike mechanical laws with long-range space-time correlations manifested as the universal inverse power law form for power spectrum concomitant with fractal geometry to the spatial pattern. Deterministic chaos in computed dynamical systems is a manifestation of quantumlike mechanical laws governing roundoff error flow dynamics with intrinsic non-local space-time connections, now identified as self-organized criticality.

## 3.3 Universal algorithm for quasicrystalline structure formation in real world and computed model dynamical systems

Observed self-organized criticality in macroscale real world dynamical systems is a signature of quantumlike mechanics implemented in unified fractal structures which coordinate the cooperative existence of fluctuations of all space-time scales in the dynamical system.

Self-organized criticality in computed model dynamical systems, also is a result of quantumlike mechanical laws with ordered growth of roundoff error structure similar to growth of large eddies from turbulent fluctuations in fluid flows. Such a concept may explain the surprising qualitative resemblance of patterns generated by computed dynamical systems to patterns in nature (Jurgen et al., 1990; Stewart, 1992) with underlying universality quantified by the Fibonacci mathematical number series.

Generation of selfsimilar patterns by space-time integration of microscopic scale fluctuations underlie observed self-organized criticality in real world and computed dynamical systems.

The universal algorithm for self-organized criticality is identified as Eq.(1), which is the law of conservation of energy for space-time fluctuations. Eq.(1) may be expressed in terms of universal Feigenbaum's constants $a$ and $d$ as (Selvam, 1993).

$$2a^2 = \pi d \qquad (3)$$

Computed solutions of disparate dynamical systems exhibit period doublings which are quantified by two universal constants $a=-2.5029$ and $d=4.6692$ named Feigenbaum's constants (Feigenbaum, 1980).

Eq.(3) states that the fractional volume intermittency of occurrence ($\pi d$) of fractal structures contributes to the total variance ($2a^2$) of the fluctuations (Selvam, 1993).

The physical mechanism underlying observed self-organized criticality in real world and computed model dynamical systems is quantified by Eq.(3), which is independent of the exact details

(physical, chemical, biological, physiological, etc.), of dynamical systems and therefore applicable to all dynamical systems.

## 3.3 Applications of the general systems theory concepts to genomic DNA base sequence structure

DNA sequences, the blueprint of all essential genetic information, are polymers consisting of two complementary strands of four types of bases: adenine (A), cytosine (C), guanine (G) and thymine (T). Among the four bases, the presence of A on one strand is always paired with T on the opposite strand, forming a "base pair" with 2 hydrogen bonds. Similarly, G and C are complementary to one another, while forming a base pair with 3 hydrogen bonds. Consequently, one may characterize AT base-pairs as weak bases and GC base-pairs as strong bases. In addition, the frequency of A(G) on a single strand is approximately equal to the frequency of T(C) on the same strand, a phenomenon that has been termed "strand symmetry" or "Chargaff's second parity". Therefore, DNA sequences can be transformed into sequences of weak W (A or T) and strong S (G or C) bases (Li and Holste, 2001). The SW mapping rule is particularly appropriate to analyze genome-wide correlations; this rule corresponds to the most fundamental partitioning of the four bases into their natural pairs in the double helix (G+C, A+T). The composition of base pairs, or GC level, is thus a strand-independent property of a DNA molecule and is related to important physico-chemical properties of the chain (Bernaola-Galvan et al., 2002). The full story of how DNA really functions is not merely what is written on the sequence of base-pairs; The DNA functions involve information transmission over many length scales ranging from a few to several hundred nanometers (Ball, 2003).

One of the major goals in DNA sequence analysis is to gain an understanding of the overall organization of the genome, in particular, to analyze the properties of the DNA string itself. Long-range correlations in DNA base sequence structure, which give rise to $1/f$ spectra have been identified (Azad et al., 2002). Such long-range correlations in space-time fluctuations is very common in nature and Li (2004) has given an extensive and informative bibliography of the observed $1/f$ noise or $1/f$ spectra, where $f$ is the frequency, in biological, physical, chemical and other dynamical systems.

The long-range correlations in nucleotide sequence could in principle be explained by the coexistence of many different length scales. The advantage of spectral analysis is to reveal patterns hidden in a direct correlation function. The quality of the $1/f$ spectra differs greatly among sequences. Different DNA sequences do not exhibit the same power spectrum. The concentration of genes is correlated with the C+G density. The spatial distribution of C+G density can be used to give an indication of the location of genes. The final goal is to eventually learn the 'genome organization principles' (Li, 1997). The coding sequences of most vertebrate genes are split into segments (exons) which are separated by noncoding intervening sequences (introns). A very small minority of human genes lack noncoding introns and are very small genes (Strachan and Read, 1996).

Li (2002) reports that spectral analysis shows that there are GC content fluctuations at different length scales in isochore (relatively homogeneous) sequences. Fluctuations of all size scales coexist in a hierarchy of domains within domains (Li et al., 2003). Li and Holste (2004) have recently identified universal $1/f$ spectra and diverse correlation structures in Guanine (G) and Cytosine (C) content of all human chromosomes.

In the following it is shown that the frequency distribution of Human chromosome Y DNA bases C+G concentration per 10bp (non-overlapping) follows the model prediction (Section 2) of self-organized criticality or quantumlike chaos implying long-range spatial correlations in the distribution of bases C+G along the DNA base sequence.

## 4. Data and Analysis

### 4.1 Data

The Human chromosome Y DNA base sequence was obtained from the entrez Databases, Homo sapiens Genome (build 34 Version 2) at http://www.ncbi.nlm.nih.gov/entrez. The ten contiguous data sets containing a minimum of 70 000 base pairs chosen for the study are given in Table 1.

Table 1: Data sets used for analyses

| Set no | Accession number | Base pairs | |
|---|---|---|---|
| | | from | to |
| 1 | NT_079581.1 | 1 | 86563 |
| 2 | NT_079582.1 | 1 | 766173 |
| 3 | NT_079583.1 | 1 | 623707 |
| 4 | NT_079584.1 | 1 | 381207 |
| 5 | NT_011896.8 | 1 | 6323261 |
| 6 | NT_011878.8 | 1 | 1089938 |
| 7 | NT_011875.10 | 1 | 9938763 |
| 8 | NT_011903.10 | 1 | 4945747 |
| 9 | NT_025975.2 | 1 | 98295 |
| 10 | NT_079585.1 | 1 | 330271 |

## 4.2 Power spectral analyses: variance and phase spectra

The number of times base C and also base G, i.e., (C+G), occur in successive blocks of 10 bases were determined in successive length sections of 70000 base pairs giving a C+G frequency distribution series of 7000 values for each data set. The power spectra of frequency distribution of C+G bases in the data sets were computed accurately by an elementary, but very powerful method of analysis developed by Jenkinson (1977) which provides a quasi-continuous form of the classical periodogram allowing systematic allocation of the total variance and degrees of freedom of the data series to logarithmically spaced elements of the frequency range (0.5, 0). The cumulative percentage contribution to total variance was computed starting from the high frequency side of the spectrum. The power spectra were plotted as cumulative percentage contribution to total variance versus the normalized standard deviation $t$. The corresponding phase spectra were computed as the cumulative percentage contribution to total rotation (Section 2.6). The statistical chi-square test (Spiegel, 1961) was applied to determine the 'goodness of fit' of variance and phase spectra with statistical normal distribution. Details of data sets and results of power spectral analyses are given in Table 2. The average variance and phase spectra for each of the ten contiguous data sets (Table 1) are given in Figure 3.

## Table 2: Results of power spectral analyses

| Set no | Base pairs used for analysis | | Number of data sets | Mean C+G concentration per 10bp | Mean $T_{50}$ | Variance spectra following normal distribution (%) | Phase spectra following normal distribution (%) |
|--------|------|--------|--------|--------|--------|--------|--------|
| | from | to | | | | | |
| 1 | 1 | 70000 | 1 | 5.47 | 6.75 | 100 | 100 |
| 2 | 1 | 700000 | 10 | 4.79 | 10.20 | 100 | 90 |
| 3 | 1 | 560000 | 8 | 4.87 | 8.71 | 100 | 100 |
| 4 | 1 | 350000 | 5 | 4.61 | 7.36 | 100 | 100 |
| 5 | 1 | 6300000 | 90 | 3.82 | 6.22 | 96.7 | 78.9 |
| 6 | 1 | 1050000 | 15 | 4.13 | 7.98 | 66.7 | 73.3 |
| 7 | 1 | 9870000 | 141 | 3.68 | 6.34 | 98.6 | 83.7 |
| 8 | 1 | 4900000 | 70 | 3.95 | 6.58 | 95.7 | 68.6 |
| 9 | 1 | 70000 | 1 | 3.89 | 3.48 | 100 | 0 |
| 10 | 1 | 280000 | 4 | 3.86 | 5.94 | 100 | 50.0 |

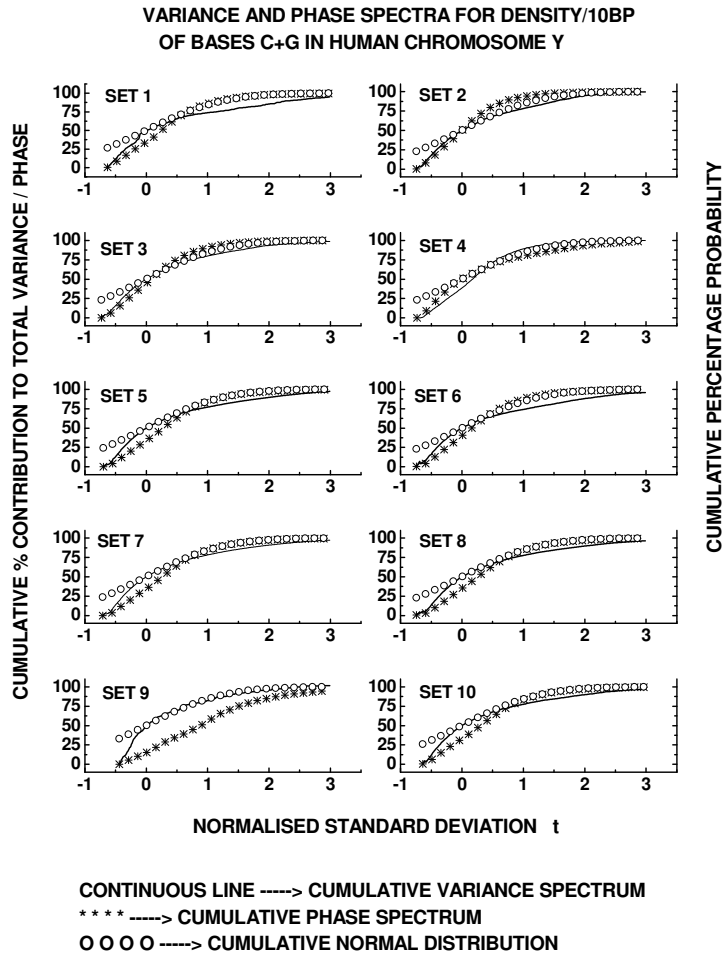

Figure 3: The average variance and phase spectra of frequency distribution of bases C+G in Human chromosome Y

## 4.3 Power spectral analyses: dominant periodicities

The general systems theory predicts the broadband power spectrum of fractal fluctuations will have embedded dominant wavebands, the bandwidth increasing with wavelength, and the wavelengths being functions of the golden mean (Section 2.4, Eq.2). The first 13 values of the model predicted (selvam, 1990; selvam and Fadnavis, 1998) dominant peak wavelengths are 2.2, 3.6, 5.8, 9.5, 15.3, 24.8, 40.1, 64.9, 105.0, 167.0, 275, 445.0 and 720 in units of the block length 10bp (base pairs) in the present study. The dominant peak wavelengths were grouped into 13 class intervals 2 - 3, 3 - 4, 4 - 6, 6 - 12, 12 - 20, 20 - 30, 30 - 50, 50 - 80, 80 − 120, 120 − 200, 200 − 300, 300 − 600, 600 - 1000 (in units of 10bp block lengths) to include the model predicted dominant peak length scales mentioned above. The class intervals increase in size progressively to accommodate model predicted increase in bandwidth associated with increasing wavelength. Average class interval-wise percentage frequencies of occurrence of dominant wavelengths are shown in Fig. 4 along with the percentage contribution to total variance in each class interval corresponding to the normalised standard deviation $t$ (Section 2.6) computed from the average $T_{50}$ (Table 2) for the ten data sets.

Figure 4: Average class interval-wise percentage distribution of dominant (normalized variance greater than 1) wavelengths is given by *line + star*. The corresponding computed percentage contribution to the total variance for each class interval is given by *line + open circle*.

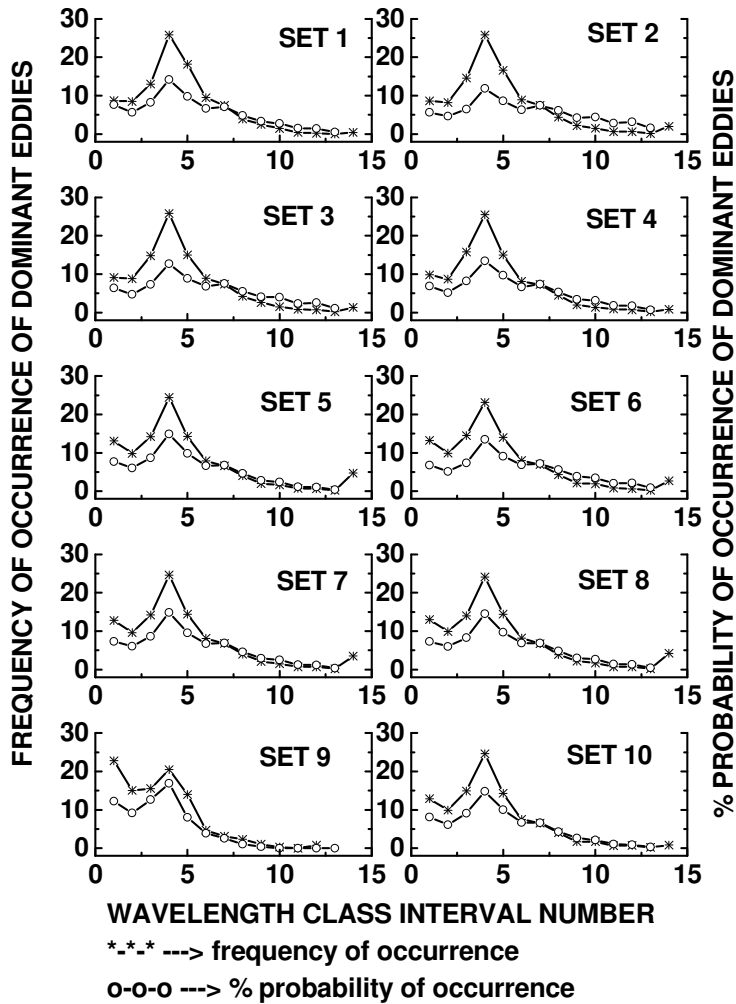

## 5. Discussions

In summary, a majority of the data sets (Table 2 and Figure 3) exhibit the model predicted quantumlike chaos for fractal fluctuations since the variance and phase spectra follow each other closely and also follow the universal inverse power law form of the statistical normal distribution signifying long-range correlations or coherence in the overall frequency distribution pattern of the bases C+G in Human chromosome Y DNA. Such non-local connections or 'memory' in the spatial pattern is a natural consequence of the model predicted Fibonacci spiral enclosing the space filling quasicrystalline structure of the quasiperiodic Penrose tiling pattern for fractal fluctuations of dynamical systems. Further, the broadband power spectra exhibit dominant wavelengths closely corresponding to the model predicted (Fig.1 and Eq.2) nested continuum of eddies. The apparently chaotic fluctuations of the frequency distribution of the bases C+G per 10bp in the Human chromosome Y DNA self-organize to form an ordered hierarchy of spirals or loops. Quasicrystalline structure of the quasiperiodic Penrose tiling pattern has maximum packing efficiency as displayed in plant phyllotaxis (Selvam, 1998) and may be the geometrical structure underlying the packing of $10^3$ to $10^5$ micrometer of DNA in a eukaryotic (higher organism) chromosome into a metaphase structure a few microns long as explained in the following. A length of DNA equal to $2\pi L$ when coiled in a loop of radius $L$ has a packing efficiency (lengthwise) equal to $2\pi L/2L = \pi$ since the linear length $2\pi L$ is now accommodated in a length equal to the diameter $2L$ of the loop. Since each stage of looping gives a packing efficiency equal to $\pi$, ten stages of such successive looping will result in a packing efficiency equal to $\pi^{10}$ approximately equal to $10^5$.

# 6. Conclusions

Real world and model dynamical systems exhibit long-range space-time correlations, i.e., coherence, recently identified as self-organized criticality. Macroscale coherent functions in biological systems develop from self-organization of microscopic scale information flow and control such as in the neural networks of the human brain and in the His-Purkinje fibers of human heart, which govern vital physiological functions.

A recently developed cell dynamical system model for turbulent flows predicts self-organized criticality as intrinsic to quantumlike mechanics governing flow dynamics. The model concepts are independent of exact details (physical, chemical, biological etc.) of the dynamical system and are universally applicable. The model is based on the simple concept that space-time integration of microscopic domain fluctuations occur on selfsimilar fractal structures and give rise to the observed space-time coherent behaviour pattern with implicit long-term memory. Selfsimilar fractal structures to the spatial pattern for dynamical systems function as dynamic memory storage device with memory recall and update at all time scales.

Overall logarithmic spiral trajectory with quasiperiodic Penrose tiling pattern are intrinsic to dynamical systems. Such spiral architecture with Fibonacci winding number and five-fold symmetry are ubiquitous in plant kingdom (Jean, 1988, 1989, 1992a,b, 1994) and are signatures of quantumlike chaos in macroscale dynamical systems. Simple laws underlie the exquisite varied patterns observed in nature.

The important conclusions of this study are as follows: (1) the frequency distribution of bases C+G per 10bp in chromosome Y DNA exhibit selfsimilar fractal fluctuations which follow the universal inverse power law form of the statistical normal distribution (Fig.3), a signature of quantumlike chaos. (2) Quantumlike chaos indicates long-range spatial correlations or 'memory' inherent to the self-organized fuzzy logic network of the quasiperiodic Penrose tiling pattern (Eq.1 and Fig.1). (3) Such non-local connections indicate that coding exons together with non-coding introns contribute to the effective functioning of the DNA molecule as a unified whole. Recent studies indicate that mutations in introns introduce adverse genetic defects (Cohen, 2002). (4) The space filling quasiperiodic Penrose tiling pattern provides maximum packing efficiency for the DNA molecule inside the chromosome.

## Acknowledgement

The author is grateful to Dr. A. S. R. Murty for encouragement.